\documentclass[12pt]{article}
\usepackage{epsfig}

\title{Gompertz law in simple computer model of ageing of biological
population}
\author{ Danuta Makowiec$^{\bf a}$, Dietrich Stauffer$^{\bf b}$ 
and Mariusz Zieli\'nski$^{\bf a}$  \\
$^{\bf a} $ Institute of Theoretical Physics and Astrophysics,\\
 Gda\'nsk University, 80-952 Gda\'nsk, ul.Wita Stwosza 57, Poland\\
$^{\bf b }$ Institute of Theoretical Physics, Cologne University,\\
D-50923 K\"oln, Euroland}

\begin{document}
\maketitle
\begin{abstract}
\noindent{
It is shown that if the computer model of biological ageing proposed by
Stauffer is modified such that the late reproduction is privileged
then the Gompertz law of exponential increase of mortality  can be
retrieved.}

\end{abstract}


\baselineskip=24pt

\section{Introduction}
Most computer simulations on biological ageing use the Penna model
\cite{model}. This model, motivated by the accumulation  mutation
hypothesis, provides a population of bit-strings that exhibits many
features known for real senescence \cite{book}. Recently Stauffer
\cite{AntiPenna} in a general review of the Penna model has suggested a
simpler alternative with less  parameters. This new attempt  is based on
the postulate of a minimum reproduction age and a maximal genetic
lifespan. Only these two numbers are transmitted from generation to
generation, with certain mutations, by asexual reproduction. 
This simple model shows the basic features required for senescence. The
age distribution shows an increase of mortality with age. 
The model reproduces successfully the catastrophic senescence of salmon
\cite{Meyer} and some specific human aspects like the social structures
of technological societies \cite{Klotz} or a transition between
extinction and survival if a minimal population for survival is required
\cite{StaufferRadomski}.

In the following we examine the other modification to the basic model.
The privilege conditions are introduced to earlier and then to late
reproduction. By doing this, in case of late reproduction,  we achieve
the exponential growth of mortality in agreement with the Gompertz law.
The increase of fertility with age exists in nature. Some fish, lobsters
and trees share this feature \cite{fertility} and therefore this life
strategy has been considered by the Penna model also \cite{book,Penna}.

\section{The model}
The population consists of $N$ individuals $i$ $(i \in 1,...,N)$
initially. Each individual is characterized by three integers: its age
$a(i)$, its minimum reproduction age $a_m(i)$ and its maximal genetic
lifespan $a_d(i)$ with $0\leq a_m(i)<a_d(i)\leq 32$.  The maximal
lifetime is restricted to 32 time units (called years), the  minimum
reproduction age may be chosen between zero and $a_d(i)-1$.  Within
these constraints the values of $a_m(i)$ and $a_d(i)$ are randomly
mutated for an offspring by $\pm1$, away from the maternal values, and
for each child separately. After an $i$th individual 
has reached its minimum reproduction age, it gives birth to one 
offspring with probability $b$, chosen as 
\begin{equation} 
b={1\over a_d(i)-a_m(i)} = { 1 \over \Delta_i }
\end{equation}
Thus during its whole mature life any individual gives one offspring at
the cumulated probability equal to 1. 

Independently of the genetic death, which happens automatically and
unavoidably if $a(i)=a_d(i)$, at each time interval an individual can
also die ``accidentally'', with the Verhulst probability $N/N_{max}$.
$N_{max}$ is called the carrying capacity to account for the fact that
any given  
environment can only support populations up to some maximal size
$N_{max}$. 
Otherwise the individuals die because of food and space limitations. 

In this paper differently than in the standard model we consider
populations in which the cumulated birth rate is not uniformly
distributed among years of reproduction: $a_m, a_m +1, \dots, a_d -1$.
Instead we propose to consider:
\begin{enumerate}
\item[(A)] privilege to younger third part of a reproduction period:
\begin{equation}
b^{young}=\cases{
2b	&	\qquad if\qquad $a_m(i)\le a(i)\le a_m(i) +{1\over 3}\Delta_i 
$ \cr
{1\over 2} b 	&\qquad 	if\qquad  $a_m(i) +{1\over 3}\Delta_i  < a(i) <
a_d(i) $   \cr}  
\end{equation}
\item[(B)] privilege to older third part of a reproduction period:
\begin{equation}
b^{old}=\cases{
{1\over 2} b	&	\qquad if\qquad $a_m(i)\le a(i)\le a_m(i) + {2\over
3}\Delta_i $ \cr
 2b 	&\qquad 	if\qquad  $a_m(i) +{2\over 3}\Delta_i  < a(i) < a_d(i) $  
\cr}  
\end{equation}
\end{enumerate}

In the following figures we present characteristics of populations
developed when three ways of reproduction are considered: standard, 
(A)  and (B) models. To avoid possible divergence we follow the Stauffer
recipe \cite{AntiPenna} and shift the birth rate accordingly:
\begin{equation}
b={1+\epsilon\over \Delta_i + \epsilon}
\end{equation}
with $\epsilon=0.08$. In figure 1 we show the population volumes in
percentage of the $N_{max}$ capacity. The population where the
reproduction of older individuals is privileged, i.e.,  model (B), is
the smallest one and varies in time by $33\%$.

Figures 2, 3 compare the distribution of $a_m$ the minimal reproduction
age and $a_d$ the maximal genetic lifespan, respectively, in the three
models considered. A significant change in the length of life of
individuals is observed when the reproduction at old age is favorable,
see figure 4 for the distribution of age $n(a)$ in a stationary
population. 

Finally figure 5 shows the mortality in the populations simulated. The
mortality is calculated according to the formula
$q(a)=1-n(a+1)\slash n(a) $ and plotted on log-scale to extract
the Gompertz law.

\section{Summary}

In Stauffer's new proposition of a model for biological ageing, by
considering the privilege conditions for time of reproduction we can
easily manipulate the mortality in a population. When reproduction at
old age is preferred then the model provides the mortality in agreement 
with the Gompertz law. 

In the simple ageing model discussed here the maximal genetic lifespan 
for each individual is limited to 32 years. This limitation was 
introduced to the model to make results better comparable with many
Penna model results. But one could ask if discarding this assumption 
would help in getting a nice Gompertz law within the standard model. 
This useful suggestion pointed out to us by the referee has motivated us for
investigations on the simple ageing model. Specially we were interested
in where the model (B) would drive a population. Preliminary studies
show that this model simulated without the limitation for the maximal
genetic lifespan leads to the stationary state such that the
distribution of $a_d$ is concentrated around age of 50. The mortality in
this population is presented in Figure 6. One can notice that the
Gompertz law of the exponential increase of mortality holds for adult
individuals at age lower than about 35.

\noindent{\bf Aknowledgement}\\
D.~M.~ thanks KBN for financial support: Project PB0273/PO3/99/16.

{\bf List of figures}

Figure 1.
Usage of environment capacity in time in standard, (A) and (B) models.

Figure 2.
Distribution of minimal reproduction age in standard, (A) and (B)
models.

Figure 3.
Distribution of lifespan in standard, (A) and (B) models.

Figure 4.
Distribution of  age in standard, (A) and (B) models.

Figure 5.
Mortality in populations of standard, (A) and (B) models.

Figure 6.
Mortality in populations of model (B) without the limitation to the
maximal genetic lifespan.

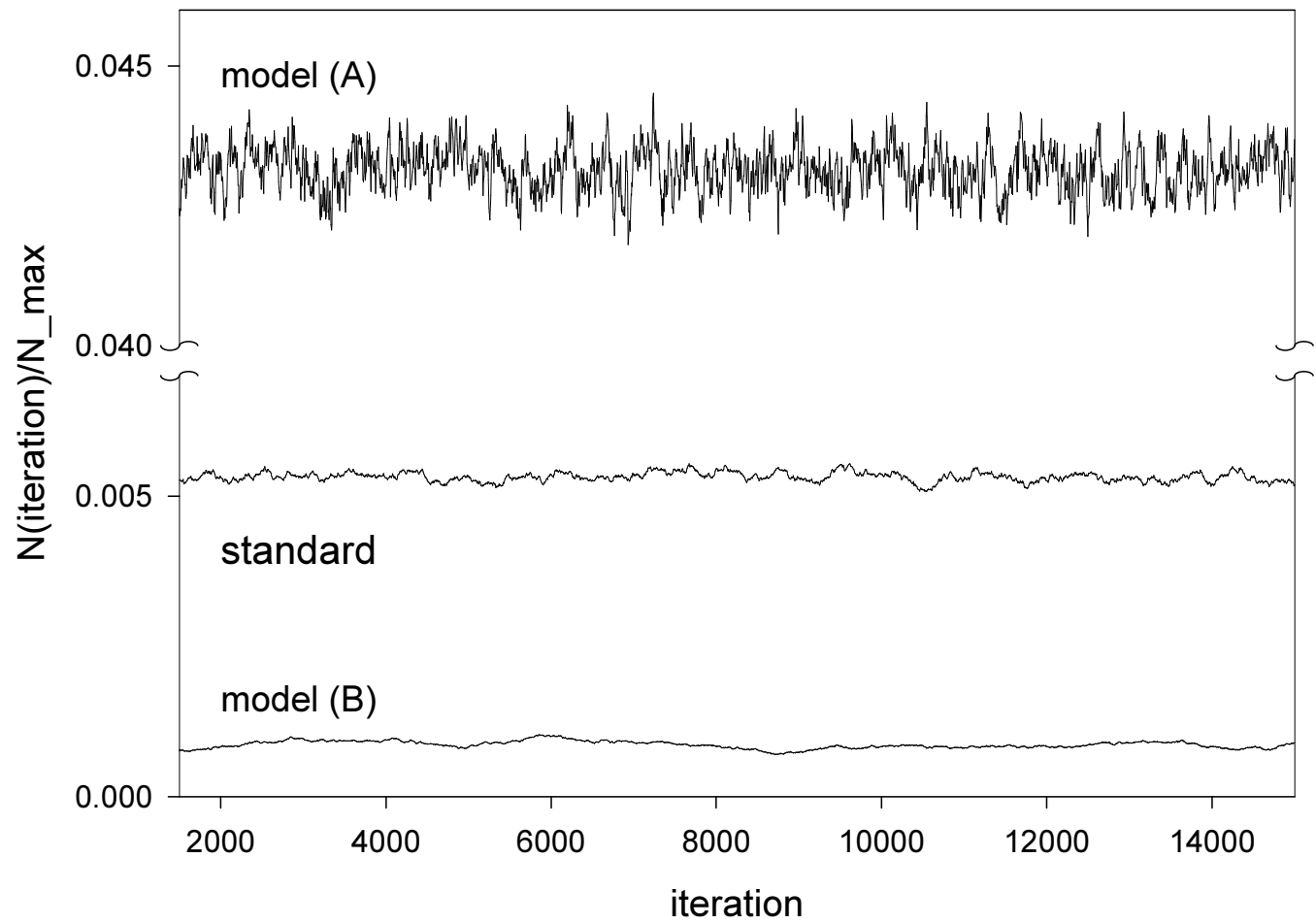
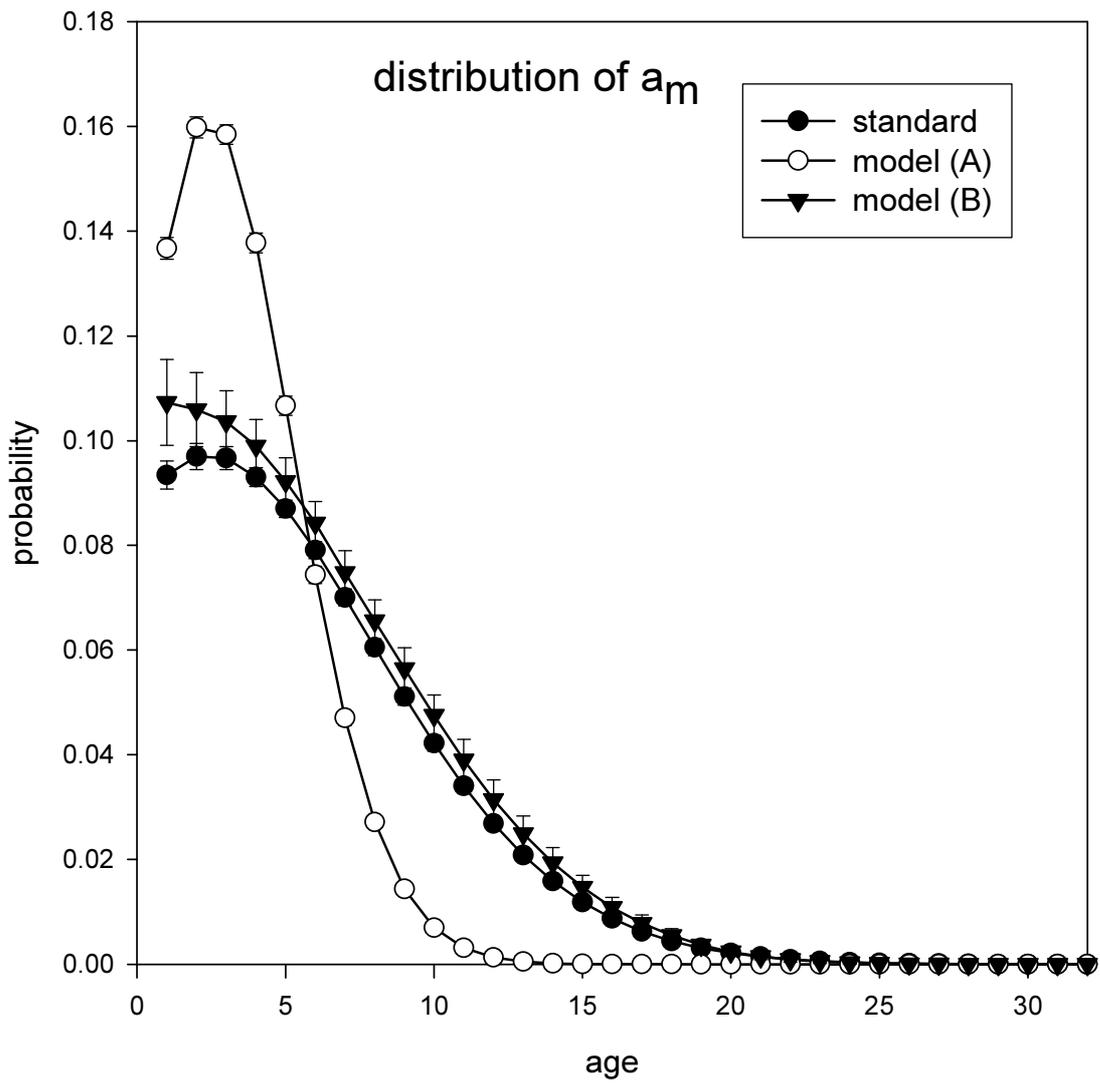
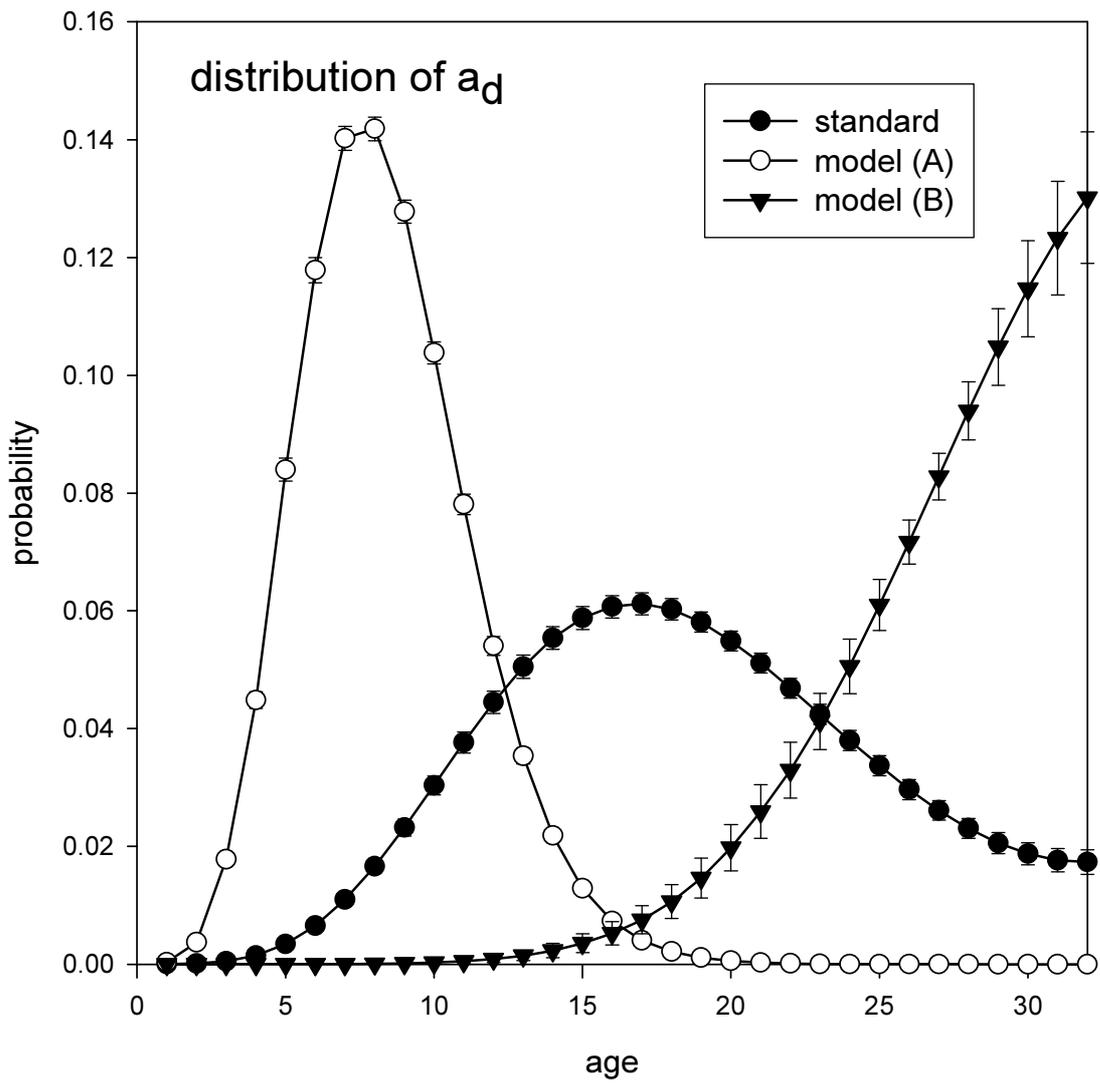
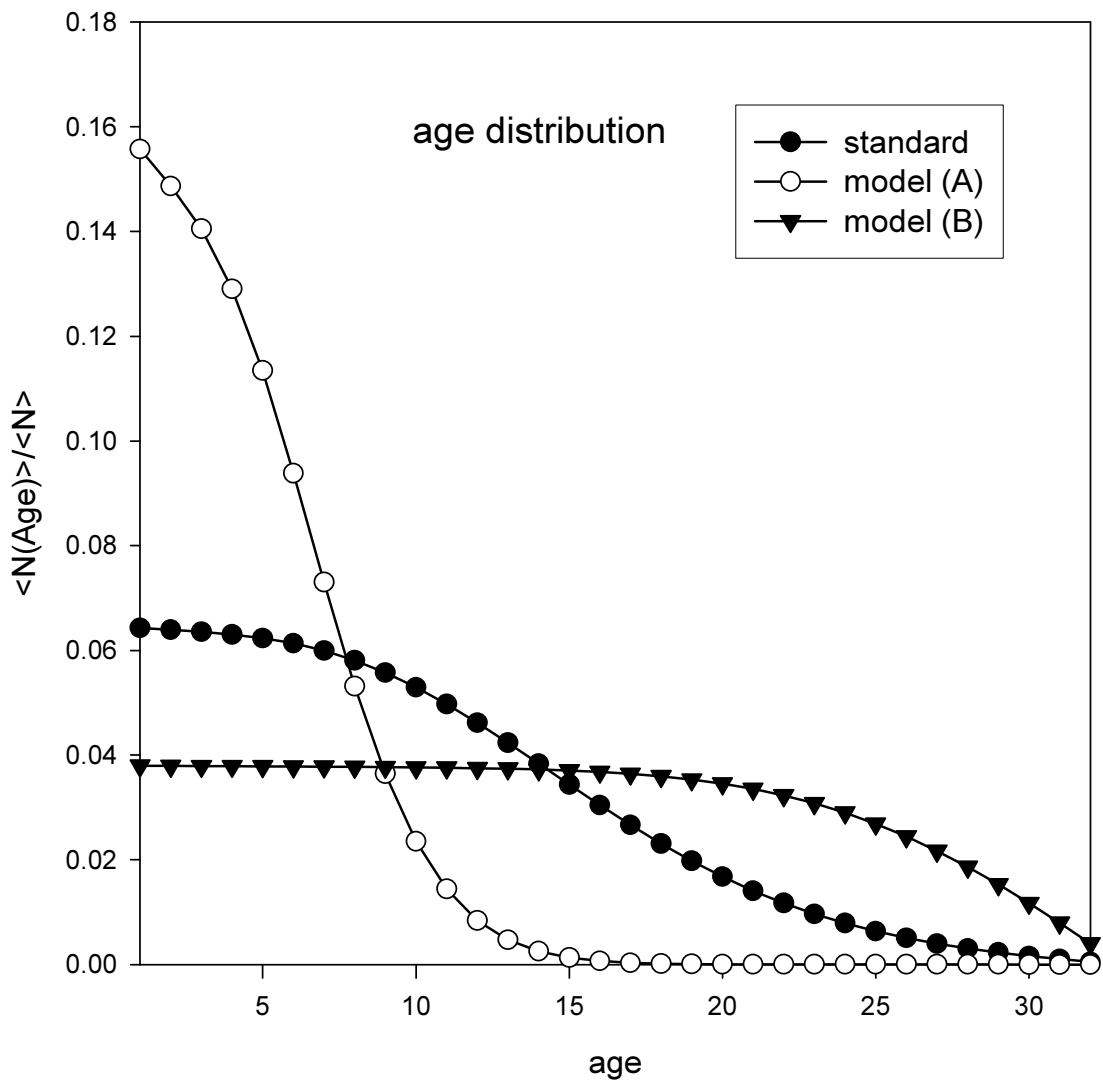
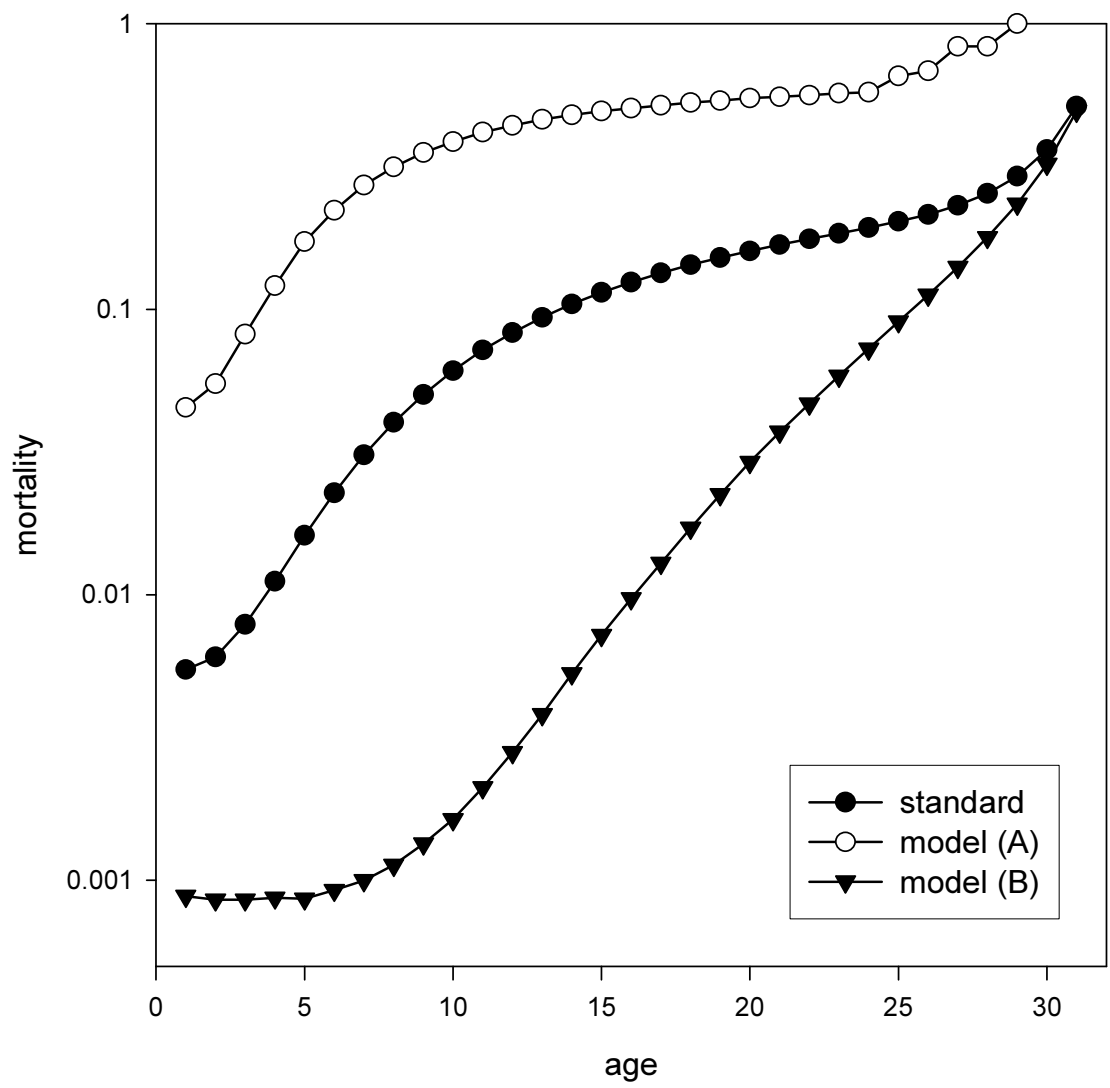
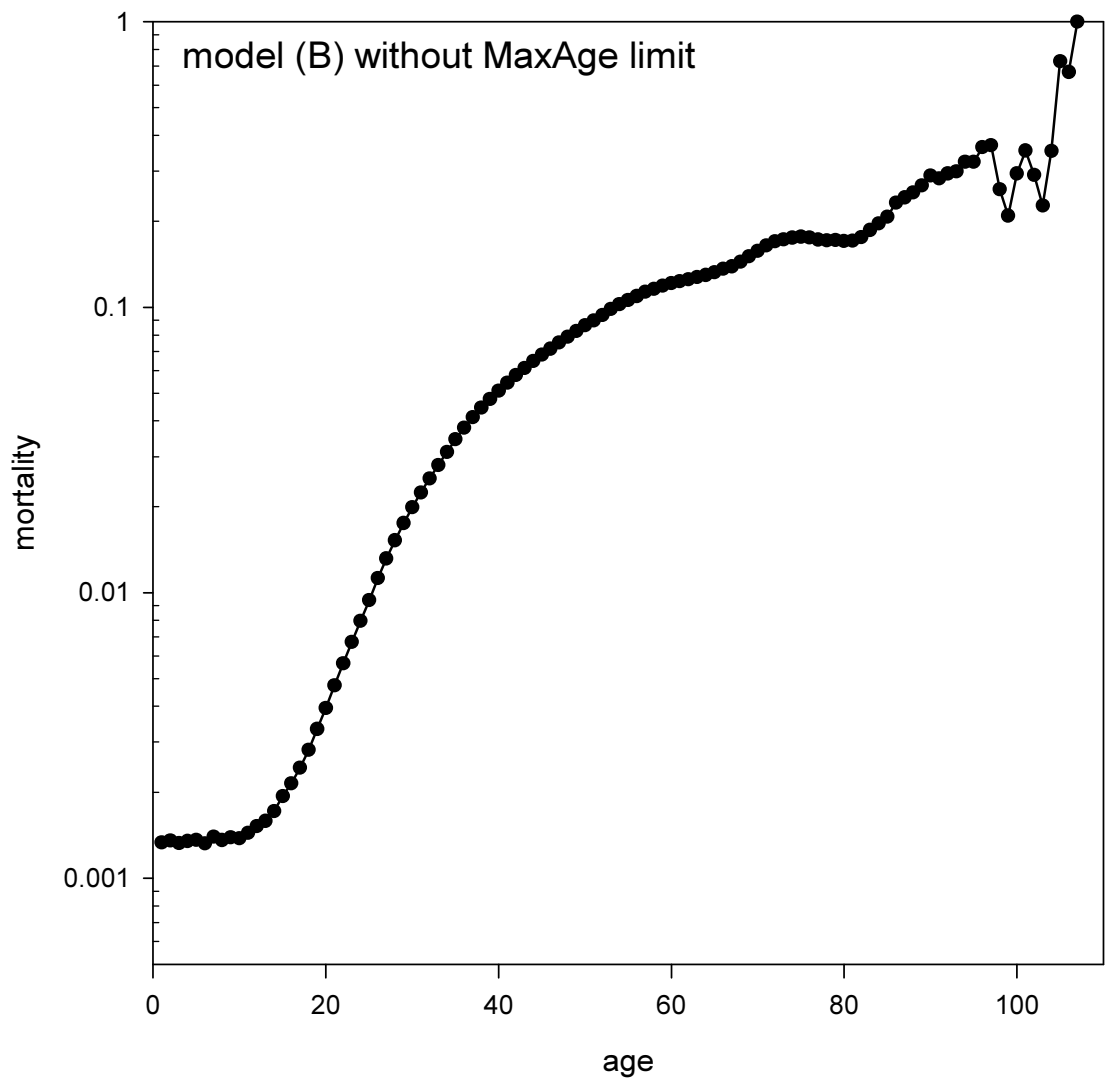
\end{document}